\documentclass[preprint,aps,showpacs,amsfonts,epsf]{revtex4}

\input{epsf.tex}

\newcommand{\E}{{\cal{E}}}
\newcommand{\s}{\sigma}

\renewcommand{\d}{{\rm d}}
\renewcommand{\a}{\alpha}

\newcommand{\be}{\begin{equation}}
\newcommand{\ee}{\end{equation}}
\newcommand{\bea}{\begin{eqnarray}}
\newcommand{\eea}{\end{eqnarray}}
\newcommand{\ba}{\begin{array}}
\newcommand{\ea}{\end{array}}
\def\J#1#2#3#4{{#1} {\bf #2}, #3 (#4)}
\def\PRD{Phys. Rev. D}
\def\PR{Phys. Rev.}

\def\APL{Ann. Phys. (Leipzig)}

\def\JMP{J. Math. Phys.}
\def\CPAM{Comm. Pure Appl. Math.}
\def\MZ{Math. Z.}
\def\TMP{Theor. Math. Phys.}

\def\CQG{Class. Quantum Grav.}

\def\MZ{Math. Zeits.}

\begin{document}
\draft
\title{The double--Reissner--Nordstr\"om solution\\ and the interaction force between\\ two
spherically symmetric charged particles}

\author{Vladimir~S.~Manko}
\address{Departamento de F\'\i sica, Centro de Investigaci\'on y de
Estudios Avanzados del IPN, A.P. 14-740, 07000 M\'exico D.F.,
Mexico}

\begin{abstract}
The physical representation of the general double--Reissner--Nordstr\"om solution is obtained by rewriting the $N=2$ Bret\'on--Manko--Aguilar electrostatic solution in the Varzugin--Chistyakov parametrization ($M_i$, $Q_i$, $R$).
A concise analytical formula is derived for the interaction force between two arbitrary Reissner--Nordstr\"om constituents, and an example of the equilibrium configuration involving two oppositely charged particles which confirms earlier Bonnor's prediction of the existence of such configurations is given. \end{abstract}


\maketitle


\section{Introduction}

The history of exact solutions of the Einstein--Maxwell equations for two aligned charged masses of the Reissner--Nordstr\"om type \cite{Rei,Nor} begins in 1917 with the two--body electrostatic solution of Weyl's class \cite{Wey} which is able to describe constituents whose masses $M_i$ and charges $Q_i$, $i=1,2$, satisfy the relation $M_1Q_2-M_2Q_1=0$. When the parameters $Q_i$ are defined by the equalities $Q_1=\pm M_1$, $Q_2=\pm M_2$, the resulting solution belongs to the Majumdar--Papapetrou family of extreme black holes \cite{Maj,Pap}, and in that case the constituents are in equilibrium which is independent of the coordinate distance $R$ separating them. The study of the general exact model characterized by arbitrary $M_i$ and $Q_i$ was started only a decade ago by Perry and Cooperstock \cite{PCo} who adjusted a known stationary axisymmetric electrovac solution \cite{MMR} for investigating the double--Reissner--Nordstr\"om equilibrium problem. The solution analyzed in \cite{PCo} was not presented in a closed analytical form since the parameters $\a_n$ it contained were implicit functions of the parameters of the axis data. A year later the multi--soliton electrostatic solution \cite{BMA} representing a system of $N$ aligned Reissner--Nordstr\"om particles was constructed with the aid of Sibgatullin's integral equation method \cite{Sib} in a concise analytical form thanks to the use of the objects $\a_n$ as arbitrary parameters. Its $N=2$ specialization, henceforth referred to as the BMA solution, fully describes the general 5--parameter double--Reissner--Nordstr\"om spacetime, the corresponding subclass of equilibrium configurations being defined by a very simple balance equation.

In a recent paper \cite{ABe} Alekseev and Belinski have been able to parametrize the 4--parameter subfamily of the BMA solution representing two Reissner--Nordstr\"om particles in equilibrium in terms of the Komar masses and charges of the constituents, the coordinate distance separating the balancing particles being a function of the mass and charge parameters. Since no technical details of the derivation have been provided in \cite{ABe}, it is one of the motivations of the present paper to demonstrate that Alekseev and Belinski's results are obtainable straightforwardly, and by purely algebraic manipulations, as a particular case of the formulas of the paper \cite{BMA} rewritten in the parametrization discovered and elaborated for the double--Reissner--Nordstr\"om problem by Varzugin and Chistyakov \cite{VCh}. For instance, for obtaining Eq.~(10) of \cite{ABe}, one only needs to rewrite the coefficient $a_{12}$ of the BMA solution (see Eqs.~(2) and (9) below) in the new parameters.

However, the removal of mystery from Alekseev and Belinski's article is not the main objective of our research. The present paper has as its principal goal the presentation of the entire 5--parameter BMA family of electrostatic solutions in the Varzugin--Chistyakov parameterization and derivation on its basis of the formula for the interaction force between two arbitrary Reissner--Nordstr\"om constituents. The extreme technical complexity of this task probably explains why neither Varzugin and Chistyakov themselves nor later on Alekseev and Belinski have been able to accomplish it. Fortunately, the difficulties of the analytic computer processing which at first glance look insuperable can be circumvented with the aid of some special tricks, but to give an idea about the scale of these difficulties, it is sufficient to mention that for instance an attempt to directly rewrite and then factorize the coefficient $a_{13}$ in terms of the new parameter set exhausts the memory of a 12~Gb RAM computer.

The paper is organized as follows. In the next section the BMA electrostatic solution in its original `$\a-\beta$' parametrization is briefly reviewed. Sec.~III is devoted to the Varzugin--Chistyakov parametrization and its relation to the parameters of the BMA solution. In Sec.~IV the general double--Reissner--Nordstr\"om spacetime is written in terms of the coordinate distance $R$ and physical parameters $M_i$ and $Q_i$, after which the derivation of the desired formula for the interaction force is carried out. Concluding remarks are given in Sec.~V.

\section{The BMA solution in the `$\a-\beta$' parametrization}

The Ernst potentials $\E$ and $\Phi$ \cite{Ern} of the BMA solution describing two aligned Reissner--Nordstr\"om particles, and the corresponding metric functions $f(\rho,z)$, $\gamma(\rho,z)$ entering the static axisymmetric line element
\be
\d s^2=f^{-1}[e^{2\gamma}(\d\rho^2+\d z^2)+\rho^2\d\varphi^2]-f\d
t^2, \label{Weyl} \ee where $\rho$ and $z$ are Weyl's cylindrical coordinates, are defined by the formulas \cite{BMA}
\bea
\E&=&\frac{A-B}{A+B}, \quad
\Phi=\frac{C}{A+B}, \quad f=\frac{A^2-B^2+C^2}{(A+B)^2}, \nonumber\\
e^{2\gamma}&=&\frac{A^2-B^2+C^2}{K_0^2 r_1r_2r_3r_4}, \quad
A=\sum\limits_{1\le i<j\le4}\a_{ij}r_ir_j, \quad
B=\sum\limits_{i=1}^4b_ir_i, \nonumber\\ C&=&
\sum\limits_{i=1}^4c_ir_i, \quad K_0=\sum\limits_{1\le
i<j\le4}\a_{ij}, \nonumber\\ \a_{ij}&=&(-1)^{i+j}(\a_i-\a_j)
\left|\begin{array}{cc} (\a_k-\beta_2)h_1(\a_k) &
(\a_l-\beta_2)h_1(\a_l)\\ (\a_k-\beta_1)h_2(\a_k) &
(\a_l-\beta_1)h_2(\a_l)\\
\end{array}\right|, \nonumber\\ &&\hspace{2cm} (i<j,k<l;l\ne i,j),
\nonumber \eea \bea b_i&=&(-1)^{i+1} \left|\begin{array}{ccc}
D(\a_k) &
D(\a_l) & D(\a_m) \\
(\a_k-\beta_2)h_1(\a_k) & (\a_l-\beta_2)h_1(\a_l) &
(\a_m-\beta_2)h_1(\a_m)\\
(\a_k-\beta_1)h_2(\a_k) & (\a_l-\beta_1)h_2(\a_l) &
(\a_m-\beta_1)h_2(\a_m)\\
\end{array}\right|, \nonumber\\
c_i&=&(-1)^{i+1} \left|\begin{array}{ccc} D(\a_k)f(\a_k) &
D(\a_l)f(\a_l) & D(\a_m)f(\a_m) \\
(\a_k-\beta_2)h_1(\a_k) & (\a_l-\beta_2)h_1(\a_l) &
(\a_m-\beta_2)h_1(\a_m)\\
(\a_k-\beta_1)h_2(\a_k) & (\a_l-\beta_1)h_2(\a_l) &
(\a_m-\beta_1)h_2(\a_m)\\
\end{array}\right|, \nonumber\\ &&\hspace{2cm} (k<l<m;k,l,m\ne i),
\nonumber\\ r_n&=&\sqrt{\rho^2+(z-\a_n)^2}.
\label{BMAsol} \eea

\noindent In the above formulas the arbitrary parameters are $\a_n$, $n=\overline{1,4}$, and $\beta_l$, $l=1,2$, which can take on arbitrary real values or occur in complex conjugate pairs, all other constant quantities, i.e., $D(\a_n)$, $f(\a_n)$ and $h_l(\a_n)$, being defined in terms of $\a_n$ and $\beta_l$ through the relations \bea D(\a_n)&\equiv&(\a_n-\beta_1)(\a_n-\beta_2), \quad f(\a_n)\equiv\frac{f_1}{\a_n-\beta_1} +\frac{f_2}{\a_n-\beta_2}, \nonumber\\ h_1(\a_n)&\equiv&e_1+2f_1f(\a_n), \quad h_2(\a_n)\equiv e_2+2f_2f(\a_n), \label{hla} \eea where \bea
e_1&=&[-2s_4+(\beta_1+\beta_2)s_3-2\beta_1\beta_2 s_2
+\beta_1^2(3\beta_2-\beta_1)s_1 \nonumber\\
&&+2\beta_1^3(\beta_1-2\beta_2)](\beta_1-\beta_2)^{-3}
-2f_1f_2(\beta_1-\beta_2)^{-1}, \nonumber\\
e_2&=&[2s_4-(\beta_1+\beta_2)s_3+2\beta_1\beta_2 s_2
-\beta_2^2(3\beta_1-\beta_2)s_1 \nonumber\\
&&-2\beta_2^3(\beta_2-2\beta_1)](\beta_1-\beta_2)^{-3}
+2f_1f_2(\beta_1-\beta_2)^{-1}, \nonumber \\
s_1&:=&\sum\limits_{i=1}^4\a_i, \quad
s_2:=\sum\limits_{1\le i<j\le4}\a_i\a_j, \quad
s_3:=\sum\limits_{1\le i<j<k\le4}\a_i\a_j\a_k, \quad
s_4:=\a_1\a_2\a_3\a_4, \label{el} \eea and \be
f_1^2=\frac{\prod_{n=1}^4(\beta_1-\a_n)}{(\beta_1-\beta_2)^2},
\quad
f_2^2=\frac{\prod_{n=1}^4(\beta_2-\a_n)}{(\beta_1-\beta_2)^2}.
\label{fl} \ee

On the upper part of the symmetry axis the potentials $\E$ and $\Phi$ behave themselves as \be \E(\rho=0,z)=1+\frac{e_1}{z-\beta_1}+\frac{e_2}{z-\beta_2}, \quad \Phi(\rho=0,z)=\frac{f_1}{z-\beta_1}+\frac{f_2}{z-\beta_2}, \label{axis_data} \ee and it is worth mentioning that $e_1$, $e_2$, $f_1$, $f_2$, together with the simple poles $\beta_1$ and $\beta_2$, can in principle be used as arbitrary parameters of the solution, but then the parameters $\a_n$ would be implicit functions of $e_l$, $f_l$, $\beta_l$ defined as roots of an algebraic quartic equation (see \cite{BMA} for details), and in such a case one would need to estimate them numerically in applications.

In view of the invariance of the field equations with respect to an arbitrary constant shift $z_0$ along the $z$--axis, the set $\{\a_n,\beta_l\}$ consisting of 6 parameters can define only 5 physical characteristics of the binary system since the constants $\a_n$ determining the location of sources on the symmetry axis (see Fig.~1) can be always set to satisfy, say, the equation $\sum\a_n=0$.

The total mass $M$ and total charge $Q$ of the system are given by the expressions \be M=-{\textstyle \frac{1}{2}}(e_1+e_2), \quad Q=f_1+f_2, \label{MQtot} \ee and these formulas play an important role in the reparametrization which will be carried out in the next section.

Although the BMA solution is analytically extended and therefore describes any combination of the black--hole and hyperextreme constituents, in what follows we shall be mainly concerned with the case of two black holes because of its greater physical importance, making where necessary only brief explanatory remarks about the binary systems involving hyperextreme constituents. In the subextreme case, all $\a_n$ are real constants to which one can assign, without loss of generality, the order \be \a_1\ge\a_2>\a_3\ge\a_4, \label{order} \ee and the locations of the upper and lower constituents are defined, respectively by the segments $\rho=0$, $\a_2\le z\le\a_1$ and $\rho=0$, $\a_4\le z\le\a_3$ of the symmetry axis (see Fig.~1). The part $\rho=0$, $\a_3< z<\a_2$ of the axis represents a Weyl strut \cite{BWe,Isr} which prevents the constituents from falling onto each other. The strut is absent in the cases when the metric function $\gamma$ is zero on the above mentioned interval separating the constituents and, as was shown in \cite{BMA}, this happens when the parameters of the BMA solution satisfy the equation $a_{12}+a_{34}=0$. The latter equation, as was already remarked in \cite{BMA2}, can be further trivialized by observing that $a_{34}=a_{12}$, so that the condition of equilibrium of two Reissner--Nordstr\"om constituents due to the balance of the gravitational and electrostatic forces is defined by vanishing of the coefficient $a_{12}$ alone: \be \a_{12}=0 \quad \Longleftrightarrow \quad (a_3-\beta_2)(a_4-\beta_1)h_1(\a_3)h_2(\a_4)-(a_3-\beta_1)(a_4-\beta_2)h_1(\a_4)h_2(\a_3)=0. \label{eq_cond} \ee

By assigning particular values to $\a_1$, $\a_2$, $\a_3$, $\a_4$, $\beta_1$ and finding numerically the corresponding value of $\beta_2$ from Eq.~(\ref{eq_cond}), the equilibrium configurations between a sub- and a hyperextreme constituent were found in \cite{BMA}, thus confirming Perry and Cooperstock's earlier results \cite{PCo} obtained with the aid of an exact solution, and Bonnor's analysis of the approximate charged two--body problem \cite{Bon}. Although the Komar masses and charges \cite{Kom} of the balancing constituents were also calculated in \cite{BMA}, it was not observed there that the `$\a$--$\beta$' parametrization is ideally appropriate for the introduction of the Komar quantities explicitly into the double--Reissner--Nordstr\"om spacetime.

\section{The Varzugin--Chistyakov parametrization}

Following the main ideas of Varzugin's study of the stationary vacuum case \cite{Var}, in the paper \cite{VCh} Varzugin and Chistyakov extended Carter's analysis of a single black hole \cite{Car} to the system of $N$ aligned charged rotating black holes supported by struts. They succeeded in relating their boundary--value problem, parametrized in terms of the individual Komar quantities and some additional parameters such as angular velocities of the horizons or electric potentials, to the matrix Riemann--Hilbert problem which they were solving on the $z$--axis. They used the properties of the reduced monodromy matrix $T(k)$ to obtain the non--linear algebraic constraints on the parameters of the boundary--value problem, thus introducing the independent parameters in terms of which all other parameters might be expressible. Varzugin and Chistyakov applied their approach to the case of two arbitrary Reissner--Nordstr\"om black holes and obtained the expressions for black holes' irreducible masses in terms of the individual Komar masses $M_i$, Komar charges $Q_i$ and the coordinate distance $R$ between the constituents (5 independent parameters in total), formulas (46) constituting the central result of the paper \cite{VCh}. As will be seen below, the knowledge of irreducible masses in terms of $M_i$, $Q_i$ and $R$ is sufficient for reparametrizing the BMA solution in terms of the latter quantities, so it will be correct to call the parameter set $\{ M_1,M_2,Q_1,Q_2,R\}$ the Varzugin--Chistyakov parametrization of the double--Reissner--Nordstr\"om problem, even though Varzugin and Chistyakov have never taken an interest in solving that problem outside the symmetry axis.

The Varzugin--Chistyakov formulas for the irreducible masses $\s_i$ written in Alekseev and Belinski's manner \cite{ABe} have the form \cite{VCh} \be \s_1=\sqrt{M_1^2-Q_1^2+2\mu\,Q_1}, \quad \s_2=\sqrt{M_2^2-Q_2^2-2\mu\,Q_2}, \quad \mu:=\frac{M_2Q_1-M_1Q_2}{M_1+M_2+R}, \label{sigmas} \ee and we refer the reader to the paper \cite{VCh} for all the details of their derivation. Note that the irreducible masses are denoted here as $\s_i$ instead of Varzugin and Chistyakov's $m_i$ because these quantities take pure imaginary values in the hyperextreme case and hence should not be confused with the genuine masses.

Formulas (\ref{sigmas}) provide us with the immediate reparametrization of the constants $\a_n$ of the BMA solution: \be \a_1={\textstyle\frac{1}{2}}R+\s_2, \quad \a_2={\textstyle\frac{1}{2}}R-\s_2, \quad \a_3=-{\textstyle\frac{1}{2}}R+\s_1, \quad \a_4=-{\textstyle\frac{1}{2}}R-\s_1, \label{an_par} \ee because, by definition, \be \s_1={\textstyle\frac{1}{2}}(\a_3-\a_4), \quad \s_2={\textstyle\frac{1}{2}}(\a_1-\a_2), \quad R={\textstyle\frac{1}{2}}(\a_1+\a_2-\a_3-\a_4), \label{si_an} \ee where $R$ is the coordinate distance between the centers of the black hole horizons (see Fig.~1). Mention that we have chosen the origin of the coordinate system in such a way that $\sum\a_n=0$, and we place the constituent denoted by index 2 above the constituent with index 1 to meet the conventions of papers \cite{VCh,ABe}.

Our next objective is to express the remaining parameters of the BMA solution, the poles $\beta_1$ and $\beta_2$, in terms of $M_i$, $Q_i$ and $R$. This can be done in the following way. First, the substitution of $e_1$ and $e_2$ from (\ref{el}) into the first formula in (\ref{MQtot}) yields the equation \be \beta_1+\beta_2={\textstyle\frac{1}{2}}s_1-M=-M, \label{eq1_b} \ee because $s_1\equiv 0$ by virtue of our choice of $\a_n$. Then, denoting the right--hand sides of the first and second equalities in (\ref{fl}) as $\omega_1$ and $\omega_2$, respectively, we pass from the second formula in (\ref{MQtot}), namely, $Q=f_1+f_2$, to the equation \be (\omega_1+\omega_2-Q^2)^2-4\omega_1\omega_2=0, \ee the left--hand side of which, after substituting into it the explicit expressions for $\a_n$, $\omega_1$, $\omega_2$, and also $\beta_2$ from (\ref{eq1_b}), factorizes into a pair of polynomials quadratic in $\beta_1$. The solution which is consistent with the second equality in (\ref{MQtot}) is \be \beta_1=-{\textstyle\frac{1}{2}}(M+D), \quad D=\sqrt{R^2+(M_1-M_2)(M_1-M_2+2R)+4Q_2(Q_1-2\mu)}, \label{b1} \ee where $M=M_1+M_2$. The expression for $\beta_2$ is now readily obtainable from (\ref{eq1_b}), yielding \be \beta_2=-{\textstyle\frac{1}{2}}(M-D). \label{b2} \ee

Once the constants $\a_n$ and $\beta_l$ are reparametrized in terms of $M_i$, $Q_i$ and $R$, no difficulty arises in rewriting the quantities $f_l$ and $e_l$ defined by formulas (\ref{fl}) and (\ref{el}) in terms of the new parameter set. Observe that from (\ref{fl}) and (\ref{MQtot}) follows that \be f_1^2-f_2^2=Q(f_1-f_2)=\omega_1-\omega_2 \quad \Longrightarrow \quad f_1-f_2=(\omega_1-\omega_2)Q^{-1}, \ee so that the expressions for $f_l$ are most easily obtainable through the formulas \be f_1={\textstyle\frac{1}{2}}[Q+(\omega_1-\omega_2)Q^{-1}], \quad f_2={\textstyle\frac{1}{2}}[Q-(\omega_1-\omega_2)Q^{-1}], \ee giving as the result \bea f_1=[QD+(M_1-M_2+R)(Q_1-Q_2)+4M_1Q_2](2D)^{-1}, \nonumber\\ f_2=[QD-(M_1-M_2+R)(Q_1-Q_2)-4M_1Q_2](2D)^{-1}, \label{fl_par} \eea where $Q=Q_1+Q_2$.

For $e_l$ we readily get from (\ref{el}): \bea e_1&=&-[M(M+D)+R(M_1-M_2)-2\mu Q]D^{-1}, \nonumber\\ e_2&=&[M(M-D)+R(M_1-M_2)-2\mu Q]D^{-1}. \label{el_par} \eea

Therefore, we have obtained all the necessary formulas for being able to rewrite the BMA solution in the Varzugin--Chistyakov parametrization.

\section{The BMA solution in the physical parametrization and the formula for the interaction force}

The simplest application of the formulas obtained in the previous section is rewriting the BMA equilibrium condition (\ref{eq_cond}) in terms of the new parameters. The substitution of (\ref{hla}), (\ref{sigmas}), (\ref{an_par}), (\ref{b1}), (\ref{fl_par}) and (\ref{el_par}) in Eq.~(\ref{eq_cond}) immediately leads to \be M_1M_2-(Q_1-\mu)(Q_2+\mu)=0, \label{bal_par} \ee and no computational problem arises during this calculation. One easily recognizes in (\ref{bal_par}) the balance equation (10) of the paper \cite{ABe}.

Turning now to the presentation of the reparametrized general BMA solution, it can be remarked that the main technical difficulty during the rewriting of the coefficients $b_i$ and $c_i$ from (\ref{BMAsol}) was finding their concise expressions which would replace the cumbersome intermediate formulas. The coefficients $a_{12}$ and $a_{34}$ which enter the expression for $A$ are the simplest ones for working out (besides, $a_{34}=a_{12}$) and are obtainable at once in their final form. On the other hand, we did not succeed in a straightforward evaluation of the coefficients $a_{13}$, $a_{14}$, $a_{23}$ and $a_{24}$; the only reasonable way of finding their reparametrized form was to process them by smaller fractions. The reparametrized BMA solution eventually takes a concise form by introducing the constant objects $\nu$ and $\kappa$ defined by \be \nu:=R^2-\s_1^2-\s_2^2+2\mu^2, \quad \kappa:=M_1M_2-(Q_1-\mu)(Q_2+\mu), \label{nk} \ee with which formulas (\ref{BMAsol}) of Sec.~II take the following elegant final form in terms of the physical parameters $M_i$, $Q_i$ and $R$: \bea
\E&=&\frac{A-B}{A+B}, \quad
\Phi=\frac{C}{A+B}, \quad f=\frac{A^2-B^2+C^2}{(A+B)^2}, \quad e^{2\gamma}=\frac{A^2-B^2+C^2} {16\s_1^2\s_2^2(\nu+2\kappa)^2 r_1r_2r_3r_4}, \nonumber \\
A&=&\s_1\s_2[\nu(r_1+r_2)(r_3+r_4)+4\kappa(r_1r_2+r_3r_4)]-(\mu^2\nu-2\kappa^2)(r_1-r_2)(r_3-r_4), \nonumber\\ B&=&2\s_1\s_2[(\nu M_1+2\kappa M_2)(r_1+r_2)+(\nu M_2+2\kappa M_1)(r_3+r_4)] \nonumber\\ &-&2\s_1[\nu\mu(Q_2+\mu)+2\kappa(RM_2+\mu Q_1-\mu^2)](r_1-r_2) \nonumber\\ &-&2\s_2[\nu\mu(Q_1-\mu)-2\kappa(RM_1-\mu Q_2-\mu^2)](r_3-r_4), \nonumber\\  C&=&2\s_1\s_2\{[\nu(Q_1-\mu) +2\kappa(Q_2+\mu)](r_1+r_2)+[\nu(Q_2+\mu)+2\kappa(Q_1-\mu)](r_3+r_4)\} \nonumber\\ &-&2\s_1[\mu\nu M_2+2\kappa(\mu M_1+RQ_2+\mu R)](r_1-r_2) \nonumber\\ &-&2\s_2[\mu\nu M_1+2\kappa(\mu M_2-RQ_1+\mu R)](r_3-r_4), \label{BMA_rep} \eea where $\s_i$ are determined by formulas (\ref{sigmas}), while the reparametrized form of the functions $r_n$ is \bea r_1&=&\sqrt{\rho^2+(z-{\textstyle\frac{1}{2}}R-\s_2)^2}, \quad r_2=\sqrt{\rho^2+(z-{\textstyle\frac{1}{2}}R+\s_2)^2}, \nonumber \\ r_3&=&\sqrt{\rho^2+(z+{\textstyle\frac{1}{2}}R-\s_1)^2}, \quad r_4=\sqrt{\rho^2+(z+{\textstyle\frac{1}{2}}R+\s_1)^2}, \label{rn_rep} \eea $r_1$ and $r_2$ referring to the upper constituent endowed with Komar quantities $M_2$, $Q_2$, and $r_3$ and $r_4$ referring to the lower constituent endowed with $M_1$, $Q_1$. Formulas (\ref{nk})--(\ref{rn_rep}) and (\ref{sigmas}) fully describe the BMA solution in the Varzugin--Chistyakov parametrization.

One can see that in the case of balancing constituents $\kappa=0$ and the expressions (\ref{BMA_rep}) simplify further. Thus, the Ernst potentials $\E$ and $\Phi$ take the form \bea \E&=&E_-/E_+, \quad \Phi=C/E_+, \nonumber\\ E_\pm&=&\s_1\s_2(r_1+r_2\pm 2M_2)(r_3+r_4\pm 2M_1) \nonumber\\ &-&[\mu(r_1-r_2)\pm 2\s_2(Q_1-\mu)][\mu(r_3-r_4)\pm 2\s_1(Q_2+\mu)], \nonumber\\ C&=&2\s_1\s_2[(Q_1-\mu)(r_1+r_2)+(Q_2+\mu)(r_3+r_4)] \nonumber\\ &-&2\mu[M_2\s_1(r_1-r_2)+M_1\s_2(r_3-r_4)], \label{potAB} \eea while for the corresponding metric functions $f$ and $\gamma$ we have \bea f&=&\frac{N}{E_+^2}, \quad e^{2\gamma}=\frac{N}{16\s_1^2\s_2^2r_1r_2r_3r_4}, \nonumber\\ N&=& [\s_1^2(r_1+r_2)^2+4\s_1^2(\mu^2-\s_2^2)-\mu^2(r_3-r_4)^2] \nonumber\\ &\times&[\s_2^2(r_3+r_4)^2+4\s_2^2(\mu^2-\s_1^2)-\mu^2(r_1-r_2)^2], \label{fAB} \eea and the expressions for $\Phi$, $f$ and $f^{-1}\exp(2\gamma)$ written in specific bipolar coordinates have already been given in \cite{ABe}. It should be emphasized that the Ernst potentials (\ref{potAB}) satisfy the field equations only when the balance condition (\ref{bal_par}) holds, i.e., when the distance $R$ is defined by the formula \be R=-M +\frac{M_2Q_1-M_1Q_2}{2(M_1M_2-Q_1Q_2)}(Q_1-Q_2\pm\sqrt{Q^2-4M_1M_2}), \label{R} \ee $M$ and $Q$ denoting, respectively, the total mass and total charge of the system. In view of this, the constant object $\mu$ entering Eqs.~(\ref{potAB}) and (\ref{fAB}) is given by the formula \be \mu=\frac{2(M_1M_2-Q_1Q_2)}{Q_1-Q_2\pm\sqrt{Q^2-4M_1M_2}}, \label{mu} \ee and the choice of the appropriate sign in the denominator of $\mu$ depends on the analysis of a concrete equilibrium position. Thus, by choosing `--' in the formula (\ref{R}), it is easy to confirm Bonnor's prediction about the possibility of the equilibrium of oppositely charged particles he made with the aid of an approximation method \cite{Bon}. An example of such equilibrium state is the following: $M_1\simeq 0.2017$, $Q_1\simeq -0.2852$, $M_2=1$, $Q_2=2$, $R\simeq 0.4688$ (the approximate numerical values are given up to four decimal places). The corresponding values for $\s_1$ and $\s_2^2$ are $\s_1\simeq 0.4409$, $\s_2^2\simeq -1.3513$, which means that the upper and lower constituents are, respectively, the hyperextreme and subextreme ones; besides, $R>\s_1$, so that the constituents do not overlap.

We now turn to the physically most important result which can be obtained with the aid of the general formulas (\ref{BMA_rep}). As was already mentioned in Sec.~II, two arbitrary Reissner--Nordstr\"om constituents are held apart by a strut when the parameters of the BMA solution do not satisfy the balance condition (\ref{eq_cond}). The analysis of the energy--momentum tensor associated with this strut permits one to introduce the interaction force between the constituents via the formula \cite{Isr,Wei} \be {\cal F} ={\textstyle\frac12}(e^{-\gamma_0}-1), \label{F} \ee where $\gamma_0$ is the constant value of the metric function $\gamma$ on the strut. In our case, formulas (\ref{F}) and (\ref{BMA_rep}) conveniently give us the following expression for ${\cal F}$: \be {\cal F}=\frac{\kappa}{\nu-2\kappa} =\frac{M_1M_2-(Q_1-\mu)(Q_2+\mu)}{R^2-(M_1+M_2)^2+(Q_1+Q_2)^2}, \label{F_BMA} \ee and it is surprising how remarkably simple is this result which is applicable to any pair of the Reissner--Nordstr\"om constituents. When $Q_1=Q_2=0$ (the pure vacuum limit), one obtains from (\ref{F_BMA}) the known expression for the interaction force between two Schwarzschild black holes \cite{Wei}: \be {\cal F}=\frac{M_1M_2}{R^2-(M_1+M_2)^2}. \label{F_Sch} \ee At large separation distances, formula (\ref{F_BMA}) gives the Newtonian expression for the force between two charged particles. It should be also pointed out that the equilibrium condition (\ref{bal_par}) is a direct consequence of the formula (\ref{F_BMA}) if one demands ${\cal F}=0$.

\section{Conclusion}

In the present paper we have succeeded in working out the physical representation of the general double--Reissner--Norstr\"om solution by rewriting the 5--parameter BMA electrostatic metric in the Varzugin--Chistyakov parametrization. Formulas (\ref{nk})--(\ref{rn_rep}), (\ref{sigmas}) contain all known exact solutions for two non--extreme Reissner--Nordstr\"om constituents and are very suitable for the use in concrete applications. The 4--parameter subclass of the BMA solution representing the balancing constituents is a special case picked out by the equilibrium condition (\ref{bal_par}), and it is worth mentioning that this subclass does not contain equatorially symmetric solutions because, apart from the Majumdar--Papapetrou extreme case, equilibrium is only possible between a black hole and a hyperextreme object. The formula for the interaction force obtained in this paper is applicable to any pair of the Reissner--Nordstr\"om constituents, which is a reflection of the analytically extended character of the BMA solution.

Since Varzugin and Chistyakov's approach to the parametrization of the multi--black--hole configurations has proved to be efficient in the electrostatic case, it would be also likely to use it, probably with some amendments, in application to the stationary axisymmetric electrovac solutions. As is known, the general system of $N$ aligned Kerr--Newman black holes is described by the Ruiz--Manko--Mart\'\i n multi--soliton electrovac metric \cite{RMM}, and we expect that at least some of its particular cases representing the two--body configurations can be worked out in the physical parametrization by establishing the relationship between the canonical parameters of the multisoliton solution and the Komar quantities introduced in the paper \cite{VCh} via the boundary Riemann--Hilbert problem.

\section*{Acknowledgments}

The author is grateful to Gennady Varzugin for interesting correspondence. Special thanks go to Erasmo G\'omez for providing the author with unlimited computer facilities without which finding a correct way of calculations would be hardly possible. This work was supported by Project 45946--F from Consejo Nacional de Ciencia y Tecnolog\'\i a, Mexico.

\newpage

\begin{figure}[htb]
\centerline{\epsfysize=120mm\epsffile{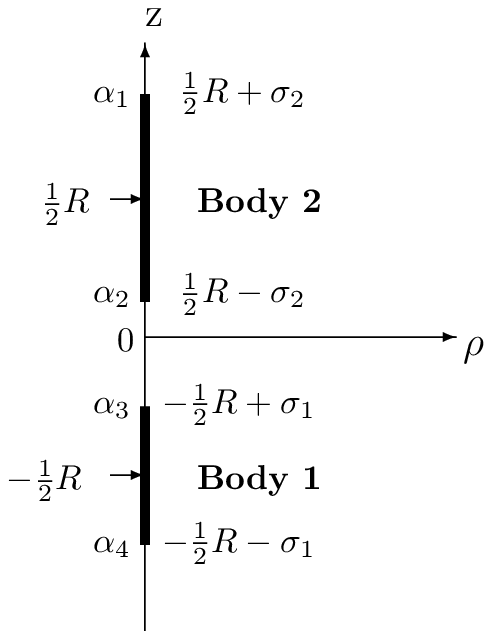}} \caption{The location of sources on the symmetry axis (the subextreme case).}
\end{figure}

\end{document}